\documentstyle[12pt]{article}
\begin{document}

\def\be{\begin{equation}}
\def\ee{\end{equation}}
\def\ba{\begin{array}{l}}
\def\ea{\end{array}}
\def\bea{\begin{eqnarray}}
\def\eea{\end{eqnarray}}
\def\no#1{{\tt   hep-th#1}}
\def\eq#1{(\ref{#1})}

\begin{flushright}
TIFR-TH-97/39 \\
hep-th/9707132
\end{flushright}
\begin{center}
\vspace*{7mm}
{\Large\bf 
Probing Type I$'$ String Theory Using D0 and D4-Branes
}\\
\vspace*{20mm}
Justin R. David$^*$, Avinash Dhar$^*$ and 
Gautam Mandal$^*$\\
Theoretical Physics Group, 
Tata Institute of Fundamental Research \\
Homi Bhabha Road, Mumbai 400 005, INDIA. \\
\vspace*{35mm}
\bf ABSTRACT\\
\end{center}
We analyse the velocity-dependent potentials seen by D0 and D4-brane
probes moving in Type I$'$ background for head-on scattering off the
fixed planes. We find that at short distances (compared to string
length) the D0-brane probe has a nontrivial moduli space metric, in
agreement with the prediction of Type I$'$ matrix model; however, at
large distances it is modified by massive open strings to a flat
metric, which is consistent with the spacetime equations of motion of
Type I$'$ theory.  We discuss the implication of this result for
the matrix model proposal for M-theory.  We also find that the
nontrivial metric at short distances in the moduli space action of the
D0-brane probe is reflected in the coefficient of the higher dimensional
$v^4$ term in the D4-brane probe action.  \vfill \hrule
\vspace*{-2mm}
\begin{flushleft}
$^*$ {\tenrm e-mail: justin, adhar, mandal,
@theory.tifr.res.in}\\ 
\end{flushleft}
\eject

\section{Introduction} 
\vspace*{7mm}

The matrix theory conjecture \cite{BFSS} is a remarkable attempt to
provide a microscopic (nonperturbative) description of M-theory
\cite{TOWN,WITTEN}, which is believed to provide a
unified description of all the known 10-dimensional superstring
theories. While evidence in favour of this conjecture has been
rapidly mounting \cite{DVV,POLPOULI,BECKER,BECKPOLTSE}, some serious
problems have also surfaced in implementing the usual prescriptions
for compactifications of M-theory within the matrix theory framework
\footnote{See \cite{BANKS} for a recent summary of developments
in this area.}.
Notable among these is the observation by Douglas, Ooguri and Shenker
\cite{DOS} that for M-theory compactifications
on curved manifolds a truncation to a finite number of
open string degrees of freedom cannot reproduce 
supergravity results. The example of K3 compactification discussed by
these authors has only eight supersymmetries. A similar problem has
also been noted \cite{GOPA} recently for possible matrix theory
descriptions of M-theory on other nontrivial backgrounds which
preserve only eight supersymmetries.

In this note we wish to point out that the matrix quantum mechanics
model constructed for Type I$'$ string theory
\cite{FERRETI,EVA,KIMREY} also suffers from this problem. Type I$'$
string theory is S-dual to $E8\times E8$ heterotic string theory and
is obtained by compactifying M-theory 
on $S^1 \times (S^1/Z_2)$ or Type
IIA on $S^1/Z^2$ \cite{HORWITT,HORWITT2}.  Vanishing of total RR
charge requires, in this compactification, the presence of 16
D8-branes.  The matrix quantum mechanics model has eight
superymmetries and belongs to the general class of supersymmetric
quantum mechanical models studied recently by Banks, Seiberg and
Silverstein \cite{SEIBERGEVA}. These authors have pointed out that in
such models the reduced supersymmetry allows for nonzero loop
corrections to the moduli space metric. In particular, in the Type
I$'$ matrix model, D0-branes perceive a nontrivial metric, first
computed in \cite{FERRETI}. This immediately raises a puzzle, as
pointed out in \cite{SEIBERGEVA}, since the spacetime equations of
motion of Type I$'$ theory lead to a flat metric and a constant
dilaton when the 8 D8-branes are placed at each of the two orientifold
planes \cite{POLWITT}. A constant dilaton \cite{SEIBERG} and a flat
metric are indeed what is perceived on a D4-brane probe. To preserve a
probe-independent notion of an underlying spacetime metric at scales
larger than the string scale one would have expected the D0-brane also
to perceive a flat metric.  Our calculations show that this is indeed
the case.  This then implies that the proposed matrix model of Type
I$'$ string theory cannot reproduce the supergravity results of the
latter.

We perform and analyse a full string theoretic computation of the
velocity-dependent potentials seen respectively by D0-brane and
D4-brane probes for head-on scattering off the fixed planes. We find
that at large distances (compared with the string scale) the
potentials seen by both the probes vanish when 8 D8-branes are on top
of the orientifold plane.  This is consistent with an underlying flat
spacetime and constant dilaton. At short distances, however, the
D0-brane has a nontrivial moduli space metric. It is clear from the
computation that massive open string excitations contribute to convert
the nontrivial metric seen at short distances to a flat metric at large
distances. Thus, there is no decoupling of massive open string degrees
of freedom as required by the matrix theory conjecture. 

The fact that at short distances the D0-brane probe sees a very
different moduli space metric than the one seen by the D4-brane
probe might seem to preclude the possibility of a probe-independent
characterization of an underlying spacetime at short 
distances. However, our analysis  of the velocity-dependent
potential seen by the D4-brane probe shows that the metric on the
moduli space of the D0-brane probe is reflected in the coefficient of
the $v^4$ term in the moduli space action of the D4-brane
probe. Thus, a probe-independent characterization of an underlying
spacetime might still exist in which the metric, dilaton etc.
simply couple differently to different probes \cite{SEIBERGEVA}.

The organization of this note is as follows.  In section \ref{sec2}
we study the head-on scattering of D0-branes and D4-branes
respectively against one of the orientifold fixed planes together
with 8
D8-branes of Type I$'$ theory.  We compute the velocity-dependent
potential seen by a D0-brane in this background and examine its
behaviour at large and small distances compared to the string
length. At large distances the dominant contribution is from 
massless closed strings.  With 8 D8-branes at each of the two
orientifold fixed planes, the entire contribution from the massless
closed strings cancels. At short distances the dominant contribution
is from the massless open string sector. For the above configuration
of D8-branes the potential starts at the $v^2$ term and gives rise to
a nontrivial moduli space metric for D0-branes. We discuss the
large $N$ limit and implications of these results for the matrix
model conjecture for M-theory.

In section \ref{sec3} we repeat the above calculation for a D4-brane
probe. At large distances we again find that the contribution from the
massless closed strings to the velocity-dependent potential
vanishes. At short distances, however, we find a nontrivial
potential. The $v^2$ term vanishes, giving rise to a flat moduli space
metric for the D4-branes, but there is a nonvanishing $v^4$ term.
Interestingly, the coefficient of this term is essentially the same as
that of the $v^2$ term in the moduli space action of the D0-brane
probe (both are proportional to $1/r^3$ where $r$ is the distance from
the orientifold plane). We discuss the implications of this for
a possible probe-independent characterization of an underlying
spacetime physics.

Section \ref{sec4} contains  a discussion of the
results and some concluding remarks.

\section{Scattering of D0-branes} \label{sec2}

Type I$'$ theory is obtained from Type IIA by compatifying the $X^9$
direction on a circle and performing a world sheet parity $\Omega$
projection and a spacetime parity $P^9$ projection along the $X^9$
direction. The $P^9$ projection leads to fixed 8-planes located at $0$
and $\pi R_{I'}$ along the $X^9$ direction, which we call the
orientifold plane or the $\Omega 8$-plane.  Each of these fixed planes
carries $-8$ units of RR charge.  Thus for charge neutrality in the
compact direction one adds 16 D8-branes. Our main  focus will be on
the configuration with 8 D8-branes at each orientifold plane. We will
probe this configuration with a D0-brane moving along the $X^9$
direction, later on to be generalized to $N$ D0-branes. In the closed
string channel the contribution to the one loop vacuum amplitude from
one of the fixed planes, say $X^9=0$, is the sum of the cylinder
diagrams between the D0-brane probe and the 8 D8-branes 
and the cross cap
between the probe and the orientifold plane. We do this
computation using the boundary state formalism for scattering
of moving D-branes \cite{CAIPOL,BIVECAN}.
\subsection{D0-$\Omega$8 interaction}
The result for the one-loop vacuum amplitude describing the static
$\Omega 8$-plane and the D0-brane moving in the $X^9$ direction is
\bea 
{\cal A }_{D0-\Omega 8} 
= 4\int_0^\infty \frac{d\tau}{\pi}
\frac{\Theta_1 '(0|i\tau + 1/2)} {i\Theta_1 (-i\epsilon |i\tau + 1/2)}
\!\times \! 
\left\{ \left[\frac{\Theta_3 (0|i\tau
+1/2)}{\Theta _2 (0|i\tau +1/2)} \right]^4 \frac{\Theta_4 (-i\epsilon
|i\tau + 1/2)}{\Theta_4 (0|i\tau +1/2)}\right. 
\\ \nonumber
- \left.
\left[\frac{\Theta_4 (0|i\tau +1/2)}{\Theta_2 (0|i\tau +1/2)}
\right]^4 \frac{\Theta_3 (-i\epsilon |i\tau +1/2)}{\Theta_3 (0|i\tau
+1/2)} \right\}  
+\infty\!\times\! 4\int_0^\infty d\tau 
\eea
This result agrees with the calculation reported recently in
\cite{LIFBER}.  In the above equation $\pi\epsilon =\tanh^{-1} v $
where $v$ is the velocity of the D0-brane along the $X^9$
direction. The infinity in the ramond sector of the amplitude is due
to the contribution of the superghost ground state, and is peculiar to
the case in which the number of Neumann-Dirichlet boundary conditions
is 8 \cite{LIFBER}.  The divergent contribution need not worry us as
it cancels in the physical amplitude which is the sum of scatterings
from the 8 D8-branes and the orientifold plane.  The sign in
front of the divergent term is fixed to be `$+$' as the D0-brane and
the $\Omega 8$-plane have opposite signs of RR charge. Note that
there is no impact parameter  for this head-on
scattering problem.  
The velocity-dependant potential ${\cal V}$ 
can be calculated from the above amplitude using the procedure of
\cite{BACHAS,DKPS}. It is given by, 
\bea
\label{formula1} 
{\cal V }_{D0-\Omega 8} 
\!=\!\frac{4\sinh
\pi\epsilon}{\sqrt{2\pi^2\alpha '}} \!\int_0^\infty\!
\frac{d\tau e^{-\frac{r^2}{2\pi\tau\alpha '}} }{\pi\sqrt\tau} 
\frac{\Theta_1 '(0|i\tau + 1/2)} {i\Theta_1 (-i\epsilon |i\tau + 1/2)}
\!\!\times \!\! 
\left\{ \!\left[\frac{\Theta_3 (0|i\tau +1/2)}
{\Theta _2 (0|i\tau +1/2)}\right]^4 \!\frac{\Theta_4 (-i\epsilon |i\tau
+ 1/2)} {\Theta_4 (0|i\tau +1/2)}\right. \\ \nonumber 
- \left. \left[\frac{\Theta_4 (0|i\tau +1/2)} {\Theta_2 (0|i\tau
+1/2)}\right]^4 \frac{\Theta_3 (-i\epsilon |i\tau +1/2)}{\Theta_3
(0|i\tau +1/2)} \right\}  
+\infty \!\times\!\frac{4\sinh
\pi\epsilon}{\sqrt{2\pi^2\alpha '}} \int_0^\infty
\frac{d\tau}{\sqrt\tau} e^{-\frac{r^2}{2\pi\tau\alpha '}} 
\eea 
where
$r^2=(X^9_{(0)} + X^0\sinh \pi\epsilon )^2$. At large distances
$(r>>\sqrt{\alpha '})$, the amplitude is dominated by
exchange of massless closed
strings. In this regime
the leading term in the potential is obtained by expanding
the $\Theta $ functions in powers of
$q=e^{-\pi\tau}$ and retaining only the
zeroth order term. We get, \be \label{formula2} {\cal V}_{D0-\Omega 8}
= \frac{r( \cosh 2\pi\epsilon - 5 )}{\pi\alpha '} \ee We have ignored
the divergent term keeping in mind that it will cancel a similar
divergent term from the 8 D8-branes.  
At short distances ($r<<\sqrt{\alpha '}$) massless open strings 
dominate the
amplitude. 
To extract their contribution
it is convenient to perform a
world sheet duality transformation and convert the cross cap to a
mobius strip. This is easily done by substituting $\tau =1/4t$ and
using the modular properties of the $\Theta$ functions. The modular
parameter of the $\Theta$ functions changes from $i\tau +1/2$ to $i/4t
+1/2$. The corresponding modular transformation is
\be 
\left(
\begin{array}{cl}
1 & -1 \\
2 & -1 
\end{array}
\right)
\ee
This can be obtained by the following sequence of S and T
transformations,
\be
TST^2S
\ee
These lead to the list of formulae given below.
\bea  \nonumber
\Theta_1 (z|i/4t +1/2) &=& \exp (i3\pi /4)(-2it)^{1/2} 
\exp(-4\pi z^2 t) \Theta_1 (2itz|it+1/2) \\ \nonumber
\Theta_1 '(0|i/4t +1/2) &=& \exp (i3\pi /4)(2it)(-2it)^{1/2} 
 \Theta_1 '(0|it+1/2) \\ \nonumber
\Theta_2 (z|i/4t +1/2) &=& \exp (i\pi /4)(-2it)^{1/2} 
\exp(-4\pi z^2 t) \Theta_2 (2itz|it+1/2) \\ \nonumber
\Theta_3 (z|i/4t +1/2) &=& \exp (i\pi /2)(-2it)^{1/2} 
\exp(-4\pi z^2 t) \Theta_4 (2itz|it+1/2) \\ \nonumber
\Theta_4 (z|i/4t +1/2) &=& (-2it)^{1/2} 
\exp(-4\pi z^2 t) \Theta_3 (2itz|it+1/2) \\ 
\eea
Using these transformations we obtain the following expression for
the potential,
\bea \label{formula3}
{\cal V}_{D0-\Omega 8}\! =\! \frac{4\sinh\pi\epsilon}
{\sqrt{2\pi^2\alpha '}}\!
\int_0^\infty\! \frac{dt e^{-\frac{2r^2t}{\pi \alpha '}} }
{\pi \sqrt{t}}
\frac{\Theta_1 '(0|it +1/2)}{\Theta_1 (2\epsilon t| it +1/2)}\!\!\times 
\!\! \left\{
\!\left[\frac{\Theta_3 (0|it +1/2)}
{\Theta_2 (0|it +1/2)}\right]^4
\!\frac{\Theta_4(\epsilon t|it +1/2)}{\Theta_4 (0|it +1/2)} \right.
\\ \nonumber
-\left. \left[\frac{\Theta_4 (0|it +1/2)}
{\Theta_2 (0|it +1/2)}\right]^4
\frac{\Theta_3 (2\epsilon t|it +1/2)}{\Theta_3 (0|it +1/2)}
\right\} 
+ \infty \!\times\!\frac{2\sinh\pi\epsilon}{\sqrt{2\pi ^2\alpha '}}
\int_0^\infty \frac{dt}{t^{3/2}}e^{-\frac{2r^2t}{\pi \alpha '}}
\eea
It is interesting to note 
an effective doubling of the distance $r$ and the rapidity
parameter $\pi \epsilon$ in the above expression.
Thus, even though \eq{formula3} 
is obtained from a closed string calculation by a world
sheet duality, the picture of an open string
stretching between the D0-brane and its image 
(reflected in the orientifold plane) 
automatically emerges.
The contribution of massless
open strings to the
potential is then given by 
\be \label{formula4}
{\cal V}_{D0-\Omega 8} = 
\frac{2\sinh\pi\epsilon}{\sqrt{2\pi^2\alpha '}}
\int_0^\infty dt
\frac{e^{\frac{-r^2t}{2\pi\alpha '}}}{\sqrt{t}\sin(\pi\epsilon t/2)}
\left(1-\frac{\cos\pi\epsilon t -1 }{4}\right)
\ee
In obtaining the above expression we have made the change of variable
$t\rightarrow t/4$. In the region $\sqrt{\alpha '}
>>r>>\sqrt{2\pi^2\alpha'\epsilon}$ we can perform a low velocity
expansion of the potential. This gives
\be \label{formula5}
{\cal V}_{D0-\Omega 8} = 
-\left( 1+\frac{v^2}{6} \right)\frac{4r}{\pi\alpha '} +      
\frac{2v^2\pi\alpha '}{3r^3} + O(v^4)
\ee
where we have used $\pi \epsilon = v + o(v^3).$
As a check on our calculations we compare 
\eq{formula2} and \eq{formula5} and note 
that the coefficient of the
leading term in the 
velocity expansion, that is the term linear in $r$,
matches. 
This is as expected since the difference in the dimension of
the probe and the target is $0$ mod $4$ \cite{DKPS}.
\subsection{D0-D8 interaction}
Let us now find the potential felt by 
a D0-brane as it scatters off the
D8-branes at the fixed plane. To do this we compute the cylinder
diagram between the D8-brane and the D0-brane. 
We find the potential to be
\bea \label{formula6}
{\cal V}_{D0-D8} =
\frac{\sinh\pi\epsilon}{\sqrt{2\pi^2\alpha '}}
\int_0^\infty d\tau
\frac{e^{\frac{-r^2}{2\pi\alpha '\tau}}}{4\sqrt{\tau}}
\frac{1}{f_2^8 (e^{-\pi \tau})}
\frac{\Theta_1 '(0|i\tau )}{i\pi\Theta_1 (-i\epsilon |i\tau)} \times
\\  \nonumber
\left\{
f_4^8(e^{-\pi \tau})
\frac{\Theta_3(-i\epsilon |i\tau )}{\Theta_3 (0|i\tau)}
-f_3^8(e^{-\pi\tau})
\frac{\Theta_4(-i\epsilon |i\tau)}{\Theta_4 (0|i\tau)}
\right\} 
-\infty\!\times\!
\frac{\sinh\pi\epsilon}{\sqrt{2\pi^2\alpha '}}
\int_0^\infty d\tau
\frac{e^{-\frac{r^2}{2\pi\alpha '\tau}}}{4\sqrt{\tau}}
\eea
which agrees with the calculations recently reported in
\cite{LIFBER,PIERRE}.
In the above expression we have fixed the sign in front of the
contribution from the Ramond
sector  to be `$-$' as the D0-brane and the D8-brane have
the same sign of the RR charge. The $f$'s are as defined in
\cite{TASI}.
We extract the contribution of the massless closed
strings relevant at long distances (neglecting the infinity as explained
before) to be
\be \label{formula7}
{\cal V}_{D0-D8} =
\frac{r( 5-\cosh 2\pi\epsilon )}{16\pi\alpha '}
\ee
To find the contribution from the 8 D8-branes at the fixed plane we
must multiply ${\cal V}_{D0-D8}$ by 16. The factor of 16 can be seen
from two points of view. The 8 D8-branes sitting at the orientifold
plane enhance the spacetime gauge group to $SO(16)$, so the
Chan-Paton factors are in the fundamental of $SO(16)\times U(1)$.
Tracing over these gives a factor of 16. If one thinks in
terms of the open string picture, we have to add
the contribution of the 8 D8-branes and their images.
This gives the factor of 16.

We note that, as  promised earlier,  in the sum
\be
{\cal V}_{D0-\Omega 8} +16{\cal V}_{D0-D8}
\ee
the infinities  cancel
leaving behind a finite answer for the physical configuration.
We also note that  this sum vanishes in
the massless closed string sector. Thus at long distances we have
a flat spacetime and constant dilaton background
from the point of view of the D0-brane probe. 

The short distance behaviour of the potential in \eq{formula6}
is found by
substituting $\tau =1/t$ and using the modular properties of the
$\Theta$ functions and the $f$'s. We get the following result
\bea \label{formula8}
{\cal V}_{D0-D8} =
\frac{\sinh\pi\epsilon}{\sqrt{2\pi^2\alpha '}}
\int_0^\infty dt
\frac{e^{\frac{-r^2t}{2\pi\alpha '}}}{4\pi\sqrt{t}}
\frac{1}{f_4^8(e^{-\pi t})}    
\frac{\Theta_1 '(0|it)}{\Theta_1 (\epsilon t|it)} \times 
\\ \nonumber
\left\{
f_2^8(e^{-\pi t})
\frac{\Theta_3 (\epsilon t|it)}{\Theta_3 (0|it)}
- f_3^8(e^{-\pi t})
\frac{\Theta_2 (\epsilon t|it)}{\Theta_2 (0|it)} 
\right\} 
-\infty\!\times\! \frac{\sinh\pi\epsilon}{\sqrt{2\pi^2\alpha '}}
\int_0^\infty
\frac{e^{-\frac{r^2t}{2\pi\alpha '}}}{4t^{3/2}}
\eea               
The contribution of the massless open string
modes  is
\be \label{formula9}
{\cal V}_{D0-D8} =
-\frac{\sinh \pi\epsilon}{\sqrt{2\pi^2\alpha '}}
\int_0^\infty dt
\frac{e^{\frac{-r^2t}{2\pi\alpha '}}}{4\sqrt{t}}
\cot \pi\epsilon t
\ee
and its velocity expansion is,
\be \label{forumula10}
{\cal V}_{D0-D8} =
\left( 1+\frac{v^2}{6} \right)
\frac{r}{4\pi\alpha '} +\frac{v^2\pi\alpha '}{12r^3}
+ O(v^4)
\ee                 
Finally, for the physical sum, we get
\be 
{\cal V}_{D0-\Omega 8} +16{\cal V}_{D0-D8}=
\frac{2v^2\pi\alpha '}{r^3} + O(v^4)
\ee
We see that, as expected, the linear 
term cancels and we are left with a
effective potential which starts off at order $v^2$. This implies
that there is a metric on the moduli space of the D0-brane. One can
write this metric as
\be 
ds^2= \frac{1}{2\lambda_{I'}}
\left( 1+ \frac{4\pi\alpha '^{3/2} \lambda_{I'}}{r^3} \right)dr^2
\ee 
where $\lambda_{I'}$ is the string coupling. 
\subsection{Large $N$ limit}

We would now like to extend the above results to the case of
scattering of a bound state of $N$ D0-branes off the fixed plane. To
the lowest order in the string coupling the potentials for this case
are simply a factor of $N$ larger than those given 
by \eq{formula1} and \eq{formula4}. This is simple to
understand for the D0-D8 scattering. For D0-$\Omega$8 scattering this
factor of $N$ arises as follows. 
To compute the mobius strip one
inserts the world sheet projection operator $\Omega$ and the space
time projection operator $P^9$ in the trace over open string states.
The eigenvalue of $\Omega P^9 $ on symmetric (antisymmetric) 
Chan-Paton factors is +1($-$1). Since
the Hamiltonian of the open string does not depend on the Chan-Paton
factors, the multiplicities come as a prefactor in the one-loop
vacuum amplitude, giving an overall factor
 $N(N+1)/2 -N(N-1)/2 =N$ in the amplitude. Thus to lowest order in
string coupling the short distance potential 
for the scattering of a bound state of N D0-branes against the
orientifold with 8 D8-branes is
\be \label{formula10.1}
{\cal V}_{D0-\Omega 8} +16{\cal V}_{D0-D8}=
\frac{N2v^2\pi\alpha '}{r^3} + O(v^4)
\ee
The nontrivial correction away from flat metric in 
\eq{formula10.1} does not disappear in the large $N$ limit even when
we take into account the $N$ scalings of space and time variables
\cite{BFSS,GOPA}.
In
higher orders of string coupling, ordinary large $N$ counting
suggests that the lowest order result could get modified by a factor
which is a function of the ratio $N/r^3$. Thus the above conclusion
cannot be modified by higher order perturbative corrections.
It is clear from our calculations that the source of the
problem is  that the massive open string modes do not
decouple at large distances. 
 The
implication for the matrix model proposal for Type I$'$ theory
\cite{FERRETI,EVA,KIMREY} is that it does not give the correct
description of the eleven dimensional physics of M-theory.

\section{Scattering of D4-branes} \label{sec3}

In order to discuss the probe-dependence of  the above results, 
we now repeat the above calculations with a D4-brane
probe. The choice of a D4-brane probe is dictated by the fact that it
is the only other probe which obeys the condition that the difference
of dimensions of the probe and the target is a multiple of four. 
\subsection{D4-$\Omega$8 interaction }
We perform similar computations as in the D0-brane case.
The velocity-dependent potential seen by the D4-brane probe due to
the orientifold is
\bea \label{formula11}
{\cal V}_{D4-\Omega 8} =
\frac{2^4(2\pi)^2\sinh\pi\epsilon V_4}{\sqrt{2\pi^2\alpha '}}
\int_0^\infty
\frac{d\tau}{\sqrt{\tau}}
\frac{e^{-\frac{r^2}{2\pi\alpha '\tau}}}{(8\pi^2\alpha ')^2}
\frac{\Theta_1 '(0|i\tau+1/2)}{i\pi\Theta_1 (-i\epsilon |i\tau +1/2)}
\times \\ \nonumber \left\{
\frac{\Theta_3 (0|i\tau +1/2)\Theta_4 (0|i\tau +1/2)} 
{\Theta_1 '(0|i\tau +1/2)\Theta_2 (0|i\tau +1/2)} 
\right\}^2 \!\times \!
\left\{
\frac{\Theta_4 (-i\epsilon |i\tau +1/2)}
{\Theta_4 (0|i\tau +1/2)}
-
\frac{\Theta_3 (-i\epsilon |i\tau +1/2)}
{\Theta_3 (0|i\tau +1/2}
\right\}
\eea
In the above expression $V_4$ is the four volume of the D4-brane. At
large distances the potential is
\be \label{formula12}
{\cal V}_{D4-\Omega 8} =
\frac{4V_4}{(8\pi^2\alpha ')^2} 
\frac{r}{\pi\alpha '}
\left(\cosh 2\pi\epsilon -1 \right)
\ee

In the variable $t=1/4\tau$ the potential is given by 
\bea \label{formula13}
{\cal V}_{D4-\Omega 8} =
\frac{2^4(2\pi)^2\sinh\pi\epsilon V_4}{\sqrt{2\pi^2\alpha '}}
\int_0^\infty
\frac{dt}{16\sqrt{t}}
\frac{e^{\frac{-2r^2t}{\pi\alpha '}}}{(8\pi^2\alpha 't)^2}
\frac{\Theta_1 '(0|it+1/2)}{\pi\Theta_1 (2\epsilon t|it +1/2)}
\times \\ \nonumber \left\{
\frac{\Theta_4 (0|it +1/2)\Theta_3 (0|it +1/2)} 
{\Theta_1 '(0|it +1/2)\Theta_2 (0|it +1/2)} 
\right\}^2 \!\times \!
\left\{
\frac{\Theta_3 (2\epsilon t |it +1/2)}
{\Theta_3 (0|it +1/2)}
-
\frac{\Theta_4 (2\epsilon t|it +1/2)}
{\Theta_4 (0|it +1/2}
\right\}
\eea
This gives the dominant contribution to the potential at short
distances to be
\be \label{formula14}
{\cal V}_{D4-\Omega 8} =
\frac{2^4 \sinh\pi\epsilon V_4}{\sqrt{2\pi^2\alpha '}}
\int_0^\infty
\frac{dt}{2\sqrt{t}}
\frac{e^{-\frac{r^2t}{2\pi\alpha '}}}{(8\pi^2\alpha 't)^2}
\frac{\cos\pi\epsilon t -1}{\sin\pi\epsilon t/2}
\ee
Note that in the above expression 
we have changed the variable of integration from $t\rightarrow
t/4$. Performing a velocity expansion, we get
\be \label{formula15}
{\cal V}_{D4-\Omega 8} =
\frac{2^4V_4}{(8\pi^2\alpha ')^2}\frac{v^2}{2}
\left(\frac{r(1-v^2/2)}{\pi\alpha '}
+\frac{v^2}{24}\frac{\pi\alpha '}{r^3} \right) + O(v^6)
\ee
\subsection{D4-D8 interaction}
The same
calculations are now repeated for the D8-brane case. The velocity
dependent potential seen by the D4-brane is
\bea \label{formula16}
{\cal V}_{D4-D8} =
\frac{V_4\sinh\pi\epsilon}{\sqrt{2\pi^2\alpha '}}
\int_0^\infty
\frac{d\tau}{4\sqrt{\tau}}
\frac{e^{\frac{-r^2}{2\pi\alpha '\tau}}}{(8\pi^2\alpha ')^2}
\frac{\Theta_1 '(0|i\tau)}{i\pi\Theta_1 (-i\epsilon|i\tau)} \times \\
\nonumber
\left\{
\frac{f_3(e^{-\pi\tau}) f_4(e^{-\pi\tau})}
{f_1(e^{-\pi\tau}) f_2(e^{-\pi\tau})}
\right\}^4 \!\times \! 
\left\{      
\frac{\Theta_3 (-i\epsilon|i\tau)}{\Theta_3 (0|i\tau)} -
\frac{\Theta_4 (-i\epsilon|i\tau)}{\Theta_4 (0|i\tau)} 
\right\}
\eea 
So at large distances we get
\be \label{formula17}
{\cal V}_{D4-D8} =
-\frac{r}{\pi\alpha '}
\frac{V_4}{(8\pi^2\alpha ')^2}
\frac{\cosh 2\pi\epsilon -1}{4}
\ee
Again we note that the sum
\be
{\cal V}_{D4-\Omega 8} + 16{\cal V}_{D4-D8}=0
\ee
At large distances this is consistent 
with the zero brane result. This confirms
that at large distances we have a probe-independent
description in terms of a flat spacetime and
constant dilaton background. 

To extract the short distance behaviour
of the potential we make the substitution $\tau =1/t$ in 
\eq{formula16}. This gives
\bea \label{formula18}
{\cal V}_{D4-D8} =
\frac{V_4\sinh\pi\epsilon}{\sqrt{2\pi^2\alpha '}}
\int_0^\infty
\frac{dt}{4\sqrt{t}}
\frac{e^{\frac{-r^2t}{2\pi\alpha '}}}{(8\pi^2\alpha 't)^2}
\frac{\Theta_1 '(0|it)}{\pi\Theta_1 (i\epsilon t|it)} \times 
\\ \nonumber
\left\{
\frac{f_3(e^{-\pi t}) f_2(e^{-\pi t})}
{f_1(e^{-\pi\tau}) f_4(e^{-\pi\tau})}
\right\}^4 \!\times\!
\left\{      
\frac{\Theta_3 (\epsilon t|it)}{\Theta_3 (0|it)} -
\frac{\Theta_2 (\epsilon t|it)}{\Theta_2 (0|it)} 
\right\}
\eea
This result has also been reported recently in
\cite{PIERRE}.
The dominant contribution at short distances is 
\be \label{formula19}
{\cal V}_{D4-D8} =
\frac{V_4\sinh\pi\epsilon}{\sqrt{2\pi^2\alpha '}}
\int_0^\infty
\frac{dt}{\sqrt t}
\frac{e^{\frac{-r^2t}{2\pi\alpha '}}}{(8\pi^2\alpha 't)^2}
\tan\pi\epsilon t/2
\ee
Making a velocity expansion, we get
\be \label{formula20}
{\cal V}_{D4-D8}=
\frac{V_4}{(8\pi^2\alpha ')^2}
\frac{v^2}{2}
\left(
-\frac{r(1-v^2/2)}{\pi\alpha '} +
\frac{v^2}{12}
\frac{\pi\alpha '}{r^3}\right) + O(v^6)
\ee
The physically relevant quantity is the sum
\be
{\cal  V}_{D4-\Omega 8} + 16{\cal V}_{D4-D8} = 
\frac{V_4}{(8\pi^2\alpha ')^2}
\frac{v^4 \pi\alpha '}{r^3} + O(v^6)
\ee
Thus the metric on the moduli space of a 
D4-brane probe is flat even at short distances. If the D8-branes were
not all placed at the orientifold plane there would be a non-trivial
moduli space metric. It turns out that this is the same as the
linear potential experienced by the D0-brane 
probe when not all the D8-branes are
placed on top of the orientifold plane.
It is  interesting  that the $1/r^3$ deviation from flat 
metric seen by the D0-brane probe is also the behaviour of
the coefficient of the higher dimensional 
$v^4$ term in the D4-brane
potential. Thus a probe-independent characterization
of an underlying spacetime may exist even at
short distances.
We also note that the leading terms in the velocity expansion of all
the individual potentials calculated in the short distance
approximation do not 
receive corrections from the
massive open string states at large distances because the
differences of the dimensions of the probes and the 
targets are always a
multiple of four \cite{DKPS}. 
But the combined potential due to the orientifold
plane and D8-branes is such that the 
leading terms always cancel.
The coefficient of the surviving next-to-leading term is
no longer an inegral
over a  modular form of weight zero and thus 
massive open string states do not decouple from 
the large distance physics.

\section{Discussion and concluding remarks} \label{sec4}

In this work we have addressed two issues. One relates to the probe
dependence of the physics of Type I$'$ background. We have seen that
at distances large compared to the string scale, the moduli space
actions for both the D0 and D4-branes are consistent with a flat
space and constant dilaton background. This probe-independence of
large distance physics is what we would have expected from previous
works \cite{POLWITT, SEIBERG} and from
physical considerations. At distances shorter than the
string scale, we get different moduli space actions for the D0 and
D4-branes. In particular, while the D0-brane perceives a nontrivial
metric, the metric seen by D4-brane is flat. 
However, this probe-dependence need not worry us. This is because
even if there exists a description of the short distance
physics in terms of an underlying spacetime picture
in a probe-independent way, the metric, dilaton, etc.
may couple differently to different probes, giving rise
to different moduli space actions. 
Thus there is no 
reason to expect the moduli space metrics to be the same
for the two probes. The good
news is that the D4-brane does have a nontrivial action. In
particular, we have seen that the coefficient of the $v^4$ term shows
exactly the same dependence on distance from the fixed plane as the
coefficient of the $v^2$ term in the D0-brane moduli space action.
This encourages us to think that even at short distances there
might be a universal notion
of an underlying spactime physics which is
simply perceived differently by different probes.

The other issue that we have addressed is the matrix theory
conjecture for M-theory in the contex of Type I$'$ compactification. 
We have argued
that the matrix model of \cite{FERRETI,EVA,KIMREY} does not reproduce
the expected spacetime gravitational physics of the Type I$'$ theory.
The massive open string modes neglected in the matrix model do not
decouple in the large distance limit. It is not clear to us  
whether any
simple modification ( which changes the number of degrees of freedom
by a finite amount ) of the matrix model would do the job.

Finally, we mention that all the systems examined so far for which the
matrix theory conjecture seems to run into problems \cite{DOS,GOPA}
have eight supersymmetries as opposed to sixteen in the model for
which the original conjecture was made\footnote{%
See, however, \cite{BERDOUG} for an example of a
quantum mechanics system with eight supersymmetries
which does reproduce supergravity results. We thank
M. Douglas for pointing out this reference to us.}.  
Matrix quantum mechanics
models with eight supersymmetries were examined in \cite{SEIBERGEVA}
and shown to admit loop corrections to the moduli space metric. In
fact, the moduli space metric is largely unconstrained.  In this
context it is interesting that the discrepancy between the moduli
space metrics for large and small black holes in the calculations of
Douglas, Polchinski and Strominger \cite{DPS} also occurs for a system
with eight supersymmetries. One possibility is that the discrepancy is
genuine and is explained by the fact that massive open string modes do
not decouple from large distance physics, just as we have seen happen
in the Type I$'$ case. In this scenario the $1/r^4$ term in the moduli
space metric, which reflects the existence of a horizon, would be
absent for small black holes. But then we must explain why
many calculations work for small black holes just as well as
for large ones\footnote{\cite{DPS}
includes a list of references on this subject.}.  
It is clearly of crucial importance to resolve these
issues for further progress in the area of black hole physics in
string theory. Another interesting example of a system
with eight supersymmetries is M-theory compactified on
$T^5/Z_2$ \cite{KESHSUN, WITT}, the matrix model construction
for which has been discussed in \cite{KIMREY2}. It would
be interesting to repeat the present analysis for this background
also.
Both these problems are currently under investigation.


\begin{thebibliography}{99}
\bibitem{BFSS} T. Banks, W. Fischler, S. H. Shenker and L. Susskind, 
``M Theory as a Matrix Model: A Conjecture, ''
 Phys. Rev. {\bf D55} (1997) 5112, \no{/9610043}.
\bibitem{TOWN} P. K. Townsend, ``The Eleven-Dimensional Membrane
Revisited,'' Phys. Lett. {\bf B350} (1995) 184,\no{/9501068}.
\bibitem{WITTEN} E. Witten, ``String Theory in Various Dimensions,''
Nucl. Phys. {\bf B443} (1995) 85, \no{/9503124}.
\bibitem{DVV} R. Dijkgraff, E. Verlinde and H. Verlinde, 
``Matrix String Theory,'' \no{/9703030}.
\bibitem{POLPOULI} J. Polchinski and P. Pouliot, ``Membrane
Scattering with M Momentum Transfer,'' \no{/9704029}.
\bibitem{BECKER} K. Becker and M. Becker, ``A Two-Loop Test 
of M(atrix) Theory,'' \no{/9705091}.
\bibitem{BECKPOLTSE} K. Becker, M. Becker, J. Polchinski and A.
Tseytlin, ``Higher Order Graviton Scattering in M(atrix) Theory,''
 \no{/9706072}.
\bibitem{BANKS} T. Banks, ``The State of Matrix Theory,''
\no{/9706168}.
\bibitem{DOS} M. R. Douglas, H. Ooguri and S. H. Shenker, ``Issues in
(M)atrix Model Compactification,'' Phys. Lett. {\bf B402} (1997) 36,
 \no{/9702203}.
\bibitem{GOPA} O. J. Ganor, R. Gopakumar and S. Ramgoolam, ``Higher
Loop Effects in M(atrix) Orbifolds,'' \no{/9705188}.
\bibitem{FERRETI} U. H. Danielsson and G. Ferretti, ``The Heterotic
Life of the D-particle,'' \no{/9610082}.
\bibitem{EVA} S. Kachru and E. Silverstein, ``On Gauge Bosons in the
Matrix Model Approach to M theory,'' Phys. Lett. {\bf B396} (1997) 76,
\no{/9612162}.
\bibitem{KIMREY} N. Kim and S. J. Rey, ``M(atrix) Theory on an
Orbifold and Twisted Membrane,'' \no{/9701139}.
\bibitem{HORWITT} P. Horava and E. Witten, ``Heterotic and Type I
String Dynamics from Eleven Dimensions,'' Nucl. Phys. {\bf B460}
(1996) 506, \no{/9510209}.
\bibitem{HORWITT2} P. Horava and E. Witten, ``Eleven Dimensional
Supergravity on a Manifold with Boundary,'' Nucl. Phys. {\bf B475}
(1996) 94, \no{/9603142}.
\bibitem{SEIBERGEVA} T. Banks, N. Seiberg and E. Silverstein, ``Zero
and One-dimensional Probes with N=8 Superysmmetry,'' Phys. Lett. {\bf
B410} (1997) 30, \no{/9703052}.
\bibitem{POLWITT} J. Polchinski and E. Witten, ``Evidence for
Heterotic-Type I String Duality,'' Nucl. Phys. {\bf B460} (1996) 525, 
\no{/9510169}.
\bibitem{SEIBERG} N. Seiberg, ``Five Dimensional SUSY Field Theories,
Non-trivial Fixed Points and String Dynamics,'' Phys. Lett {\bf B388}
1996 753, \no{/9608111}.
\bibitem{CAIPOL} Y. Cai and J. Polchinski, ``Consistency of Open
Superstring Theories,'' Nucl. Phys. {\bf B296} (1988) 91.
\bibitem{BIVECAN} M. Billo, P. Di Vecchia and  D. Cangemi, ``Boundary
States for Moving D-branes,'' \no{/9701190}.
\bibitem{LIFBER} O. Bergman, M. R. Gaberdiel and G. Lifschytz,
``Branes,  Orientifolds and the Creation of Elementary Strings,''
\no{/9705130}.
\bibitem{PIERRE} J. M. Pierre, ``Interactions of
Eight-branes in String Theory and M(atrix) Theory,'' \no{/9705110}.
\bibitem{BACHAS} C. Bachas, ``D-Brane Dynamics,'' Phys. Lett. {\bf
B374} (1996) 37, \no{/9511043}.
\bibitem{DKPS} M. R. Douglas, D. Kabat, P. Pouliot and S. H. Shenker,
``D-branes and Short Distances in String Theory,'' Nucl. Phys. {\bf
B485} (1997) 85,  \no{/9608024}.
\bibitem{TASI} J. Polchinski, ``TASI Lectures on D-branes,''
\no{/9611050}.
\bibitem{BERDOUG} M. Berkooz and M. R. Douglas, ``Five-branes
in M(atrix) Theory,'' Phys.Lett. {\bf B395} (1997) 196,
\no{/9610236}. 
\bibitem{DPS} M. R. Douglas, J. Polchinski and A. Strominger,
``Probing Five-Dimensional Black Holes with D-branes,'' \no{/9703031}
\bibitem{KESHSUN} K. Dasgupta and S. Mukhi, ``Orbifolds of
 M-theory,'' Nucl. Phys. {\bf B465} (1996) 399, \no{/9512196}.
\bibitem{WITT} E. Witten, ``Five-branes And $M$-Theory On An 
Orbifold,'' Nucl.Phys. {\bf B463} (1996) 383, \no{/9512219}.
\bibitem{KIMREY2} N. Kim and S. J. Rey, ``M(atrix) Theory on
$T_5/Z_2$ Orbifold and Five-Branes,'' \no{/9705132}.
\end{thebibliography}
\end{document}